\documentclass[11pt]{article}
\usepackage{amsmath}
\usepackage{enumerate}
\usepackage{rotating} 
\usepackage{color}
\usepackage{mathtools}
\usepackage{url}
\usepackage{booktabs}
\usepackage{natbib, multirow}
\usepackage{amsmath,amssymb}

\usepackage{nicefrac}
\usepackage{appendix}
\usepackage{bm}  

\usepackage{accents}
\newcommand{\ubar}[1]{\underaccent{\bar}{#1}}

\definecolor{Dblue}{rgb}{0.1,0,0.55}
\definecolor{mygreen}{rgb}{0.1,0.6,0.2}
\definecolor{myorange}{rgb}{0.96, 0.49, 0.26}
\definecolor{turquoise}{rgb}{0.19, 0.84, 0.78}

\usepackage[margin=1in]{geometry}

\def\bLambda{\mbox{\boldmath $\Lambda$}}
\def\brho{\boldsymbol{\rho}}

\def\bb{\mbox{\boldmath $b$}}

\def\b0{\mbox{\bf{0}}}

\def\bone{\mathbf{1}}
\def\bC{\mathbf{C}}
\def\bK{\mathbf{K}}
\def\bR{\mathbf{R}}
\def\bV{\mathbf{V}}
\def\bW{\mathbf{W}}
\def\bx{\mathbf{x}}
\def\by{\mathbf{y}}

\usepackage{float}
\usepackage{setspace}
\renewcommand\l{\left}
\renewcommand\r{\right}

\begin{document}

\singlespacing
\title{\Large \textbf{Prediction Using a Bayesian
Heteroscedastic Composite Gaussian Process}}
\author{Casey B.~Davis$^{a}$, Christopher M.~Hans$^{b}$, Thomas J.~Santner$^{c}$ \\
\small \emph{Department of Statistics, The Ohio State University, Columbus, OH 43210, USA}\\
\footnotesize $^a$cbdavis33@gmail.com; $^b$hans@stat.osu.edu; $^c$santner.1@osu.edu}

\date{}
\maketitle


\doublespacing

\begin{abstract}

This research proposes a flexible Bayesian extension of the 
composite Gaussian process (CGP) model of Ba and Joseph (2012)
for  predicting (stationary or) non-stationary $y(\bm{x})$. 
The CGP generalizes the regression plus 
stationary Gaussian process (GP) model by replacing the regression 
term with a GP.  
The new model, $Y(\bm{x})$, 
can accommodate large-scale trends estimated by 
a global GP, local trends estimated by an independent local GP, and
a third process to describe 
heteroscedastic data in which $Var(Y(\bm{x}))$ 
can depend on the inputs.
This paper proposes a prior which ensures
that the fitted 
global mean is smoother than the local deviations,
and extends the covariance structure of the CGP to allow
for differentially-weighted global and local components.
A Markov chain 
Monte Carlo algorithm is proposed to provide
posterior estimates of the
parameters, including the values
of the heteroscedastic variance at the training and test data
locations.  
The posterior distribution is used to
make predictions and to quantify the 
uncertainty of the predictions using
prediction intervals.  The method is illustrated using both
stationary and non-stationary $y(\bm{x})$.

\end{abstract}

\vspace{-.15in}



\section{Introduction}
\label{sec:intro}

We introduce a Bayesian composite Gaussian process as a
model for generating and predicting 
non-stationary functions $y(\bm{x})$ 
defined over an
input space $\mathcal{X}$.  Our model is motivated by and extends the work
of \citet{ba:12}, who introduced a composite Gaussian process (CGP) 
as a flexible model for  $y(\bm{x})$.  They used 
 $y(\bm{x})$ evaluations at training data locations 
$\bx_i$, $i = 1, \ldots, n$,  to predict $y(\bm{x})$ at one or more 
new locations and to 
quantify uncertainty about these predictions.  

The problem of predicting functions $y(\bm{x})$ that are
possibly non-stationary is particularly relevant,
as many physics-based and other simulator models 
have been developed as alternatives to physical 
experimental platforms.
Termed ``computer experiments'', 
simulator-based studies have been used, for example,  
to determine the 
engineering design of aircraft, automobiles, and prosthetic
devices, to optimize the manufacturing settings of 
precision products by injection molding,
and to evaluate public policy options 
\citep{OngSanBar2008, VilCheRac2017, LemSchBan2000}.

A common approach to prediction and uncertainty quantification when 
analyzing data from a computer experiment
is to represent the unknown function $y(\bm{x})$ 
as a realization of a Gaussian process (GP).
As there are many possible functions that are consistent with the observed values 
$y(\bx_i)$ sampled at training locations $\bx_i$, a GP is used
as a prior distribution over an infinite-dimensional space of functions.  When combined 
with the observed data, the resulting posterior distribution over functions can be used for 
prediction and uncertainty quantification.  The use of a GP as a prior over functions 
was introduced by \citet{ohag:78} in a Bayesian regression context. This
approach has subsequently been extended and used extensively in various settings related
to both physical and computer experiments 
\citep[e.g.,][]{sack:89, neal:98, kenn:01, oakl:02, bane:04, oakl:04,
SanWilNot2018}. 

Our interest lies in prediction and uncertainty quantification for functions that, when viewed
as a draw from a GP, exhibit features inconsistent with stationarity, i.e.~where the behavior 
of the function can be substantially different in different regions of the input space.
Several existing methodologies exist for working with data generated by such functions.  Perhaps
the most widely-used is universal kriging \citep[][]{cres:93}, which assumes the function
$y(\bm{x})$ can be viewed as a draw from a GP of the form 
\begin{align}
	Y(\bm{x}) = \sum_{j=1}^p f_j(\bm{x})\beta_j + Z(\bm{x}) = 
		\bm{f}^\top(\bm{x})\boldsymbol\beta + Z(\bm{x}),
		\label{eq:ukrig}
\end{align}
where $\bm{f}(\bm{x}) 
= 
(f_1(\bm{x}),\ldots,f_p(\bm{x}))^\top$ 
is a vector of known regression 
functions, $\boldsymbol\beta = (\beta_1, \ldots, \beta_p)^\top$ is a 
vector of unknown regression coefficients, and 
$Z(\bm{x})$
is a  stationary Gaussian process with mean zero,
process variance $\sigma_Z^2$,
and (positive definite) correlation function
$R(\bm{\cdot})$ so that $Z(\bm{x})$
has covariance 
\[
	\mbox{Cov}(Z(\bm{x}),Z(\bm{x + h})) = \sigma^2_Z R(\bm{h}) .
\]
Throughout this paper the notation $Z(\bm{x}) \sim 
\mbox{GP}(0,\sigma^2_Z,R(\bm{\cdot}))$ will be used to
describe this stationary process assumption.  

The intuition of the model is that
$E(Y(\bm{x})) = 
\bm{f}^\top(\bm{x})\boldsymbol\beta$ 
describes large-scale 
$y(\bm{x})$
trends while $Z(\bm{x})$ describes small-scale deviations from
the large-scale behavior.
A special case of universal kriging is ordinary kriging 
which assumes $Y(\bm{x})$ has constant mean.
\citet{cres:93} and \citet{SanWilNot2018}
provide details about the model (\ref{eq:ukrig}), 
including parametric options for $R(\bm{h})$,
methods for estimating model parameters, 
prediction methodology for test data inputs, 
and uncertainty quantification
of the predictions.

While universal kriging has proved useful in many applications, 
several limitations
have been identified. The requirement that the regression 
functions be known or 
adaptively selected from a pre-defined collection of regression functions
sometimes proves difficult.
In addition to bias due to potential misspecification of the
regression functions, standard prediction intervals under universal kriging do not
account for uncertainty in the selection of the regression functions.  From a computational
perspective, entertaining a large class of potential regression functions may 
result in a large selection problem, necessitating a combinatorial search over a large 
space.  Finally, the kriging methods described above are based on trend-stationary Gaussian
processes.  In many applications, even if the mean function is appropriate, the unknown function 
being emulated may exhibit non-stationary behavior due to the variance function.  Ignoring
these aspects of the data may result in both poor prediction and inaccurate
 uncertainty quantification.

\begin{figure}[t] 
  \begin{center} 
	\includegraphics[width = 0.9\linewidth]{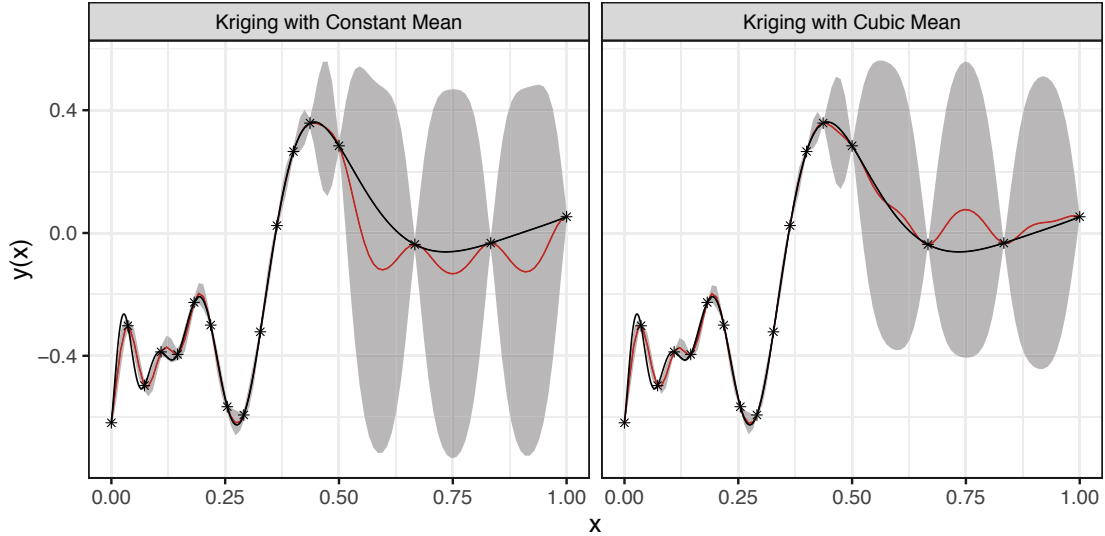}
	\caption{\small \sl  Kriging predictors (red lines)
		for the BJX function (black lines) given in equation~(\ref{eq:bjx}) 
		based on the training data shown as black points together with
		95\% prediction intervals.  Left Panel:
		constant mean; Right Panel: cubic mean. 		
	\label{fig:bjxMPERK}}
  \end{center}
\end{figure}

As a motivating example, consider the (non-stationary) function 
\begin{align}
	y(x) = \sin\left( 30(x-0.9)^4 \right) \cos\left( 2(x-0.9) \right) + \frac{(x-0.9)}{2}, \;\;x \in [0,1], 
		\label{eq:bjx}
\end{align}
which was originally considered by \citet{xion:07} and also by \citet{ba:12} (we henceforth
refer to (\ref{eq:bjx}) as the BJX function). Figure~\ref{fig:bjxMPERK} plots
the BJX function as a black line.  
The points in the figure indicate the value of the 
function at the
$n = 17$ training data locations used by \citet{ba:12}. 
If viewed as a realization of a 
stochastic process, one might describe the BJX 
function as having three behavior paradigms. 
For small $x$, $y(x)$ can be described as having
a relatively flat global trend with rapidly-changing local adjustments.  For 
intermediate $x$, $y(x)$ increases rapidly and smoothly, 
with few local departures.
For large $x$, $y(x)$ has a relatively flat global trend with 
minor local adjustments.

Two aspects of universal kriging (UK) prediction of the BJX function are of interest: the accuracy 
of the point predictions and the narrowness of the associated uncertainty band. Figure~\ref{fig:bjxMPERK} 
shows point predictions of $y(x)$ for the constant- and cubic-mean UK predictors computed at a 0.01 
grid of prediction locations; a nugget was not included and so the predictors interpolate at the 17 training 
data locations. While the constant- and cubic-mean predictors and uncertainty bands are similar for $x < 0.5$, 
differences can be seen when $x > 0.5$.  Reversion to the global mean is evident for the constant-mean 
predictor, while the cubic-mean predictor exhibits a ``bump'' near $x = 0.75$ that is driven by reversion
to the estimated cubic mean function. The 95\% prediction intervals based on the cubic mean are shorter than 
those based on the constant mean, however both sets of intervals are unreasonably wide when $x > 0.5$.  
Intuitively, the intervals should be short where y(x) is essentially flat.

To address these shortcomings, alternatives to universal kriging have been 
proposed. The treed Gaussian processes (TGPs) of \citet{gram:08} are one such alternative. 
The TGP model assumes that the input space can be partitioned into  
rectangular subregions so that a GP with a linear trend and stationary covariance structure 
is appropriate to describe $y(\bm{x})$ in each region. Following \citet{brei:84}, TGP methodology 
partitions the input space
by making binary splits on a sequence
of the input variables over their ranges,
where splits can be made on previously-split inputs by
using a subregion of the previous range.
After the 
input space is partitioned, 
the data in each region are used to fit a prediction model independently 
of the fits for other regions. In earlier proposals for
fitting data to each region, \citet{brei:84} 
 fit a constant mean model to the data in each region, and
\citet{chip:98} fit a Bayesian hierarchical 
linear model in each region. 
The TGP model extends \citet{chip:98} by fitting a 
GP with a linear trend and stationary 
covariance structure in each region. 
While TGP prediction can have computational advantages over kriging,
one disadvantage is that the method can be numerically challenged
when the number of training data locations in one or more regions is
small, a situation often encountered in computer experiments.

\citet{ba:12} provide another alternative to 
universal kriging
for emulating functions exhibiting non-stationary behavior. Their composite Gaussian process (CGP) 
avoids specification and/or selection of regression functions 
that might be required to generate the unknown $y(\bm{x})$ by 
specifying the generating GP $Y(\bm{x})$
as the sum 
\[
	Y(\bm{x}) = Z_{G}(\bm{x}) 
	+ 
	\sigma(\bm{x}) Z_{L}(\bm{x}),
\]
where, conditionally on model parameters $\bLambda_{\textsc{cgp}}$,
$Z_G(\bm{x})$ and $Z_L(\bm{x})$ are independent GPs such that
$$
Z_{G}(\bm{x}) \mid \bLambda_{\textsc{cgp}} \sim \mathrm{GP}(\beta_{0}, \sigma_G^2,  G(\cdot)) 
\ \ \ \mbox{and}
\ \ \	Z_{L}(\bm{x}) \mid \bLambda_{\textsc{cgp}}  \sim \mathrm{GP}(0, 1, L(\cdot)).
$$
Under this specification, 
$Z_{G}(x)$ represents a smooth process
that captures any global trend in $y(\bm{x})$,
and 
$Z_{L}(\bm{x})$ represents
a less-smooth
process that introduces local adjustments to capture 
the function $y(\bm{x})$.
By replacing the regression term in (\ref{eq:ukrig}) 
with the more flexible $Z_{G}(\bm{x})$ process,
the CGP model $Y(\bm{x})$ is able to adapt to large-scale global 
features of $y(\bm{x})$.

 \citet{ba:12} 
employ
Gaussian correlation functions $G(\bm{h} \mid \brho_G) = \prod_{j=1}^d \rho_{G,j}^{h_j^2}$ and
$L(\bm{h} \mid \brho_L) 
= \prod_{j=1}^d \rho_{L,j}^{h_j^2}$ for the global and local processes,
respectively, where $\brho_G = (\rho_{G,1}, \ldots, \rho_{G,d})^\top$ and 
$\brho_L = (\rho_{L,1}, \ldots, \rho_{L,d})^\top$ are corresponding correlation parameters.
To ensure that the global process is smoother than the local process---and hence is interpretable as
a global trend---a vector of positive bounds $\bb$ is specified so that
$0 \leq \rho_{L,j} \leq b_j \leq \rho_{G,j} \leq 1$, $j = 1, \ldots, d$. 
Even though the conditional process mean, 
$E(Y(\bm{x}) \mid \bLambda_{\textsc{cgp}}) = \beta_{0}$, is  
constant across the input
space, the examples in \citet{ba:12} 
and that of the BJX function below
show that CGP often has greater prediction accuracy 
than ordinary kriging or even universal kriging
(when the global trend is difficult to capture with 
pre-specified regression functions).

The variance of the CGP $Y(\bm{x})$ is 
$\mbox{Var}( Y(\bm{x}) \mid \bLambda_{\textsc{cgp}}) 
= 
\sigma_G^2 + \sigma^2(\bm{x})$.
The term $\sigma(\bm{x})$ is a positive function that 
allows the range of the local process $Y_L(\bm{x})$, and hence the range of 
$Y(\bm{x})$, to vary 
over the input space. 
\citet{ba:12} describe an algorithm for estimating $\sigma^2(\bm{x})$ that is
implemented in their 
\texttt{R} package \texttt{CGP} \citep{ba:18}.

\begin{figure}[h]
 \begin{center}
  \includegraphics[width = 0.5\linewidth]{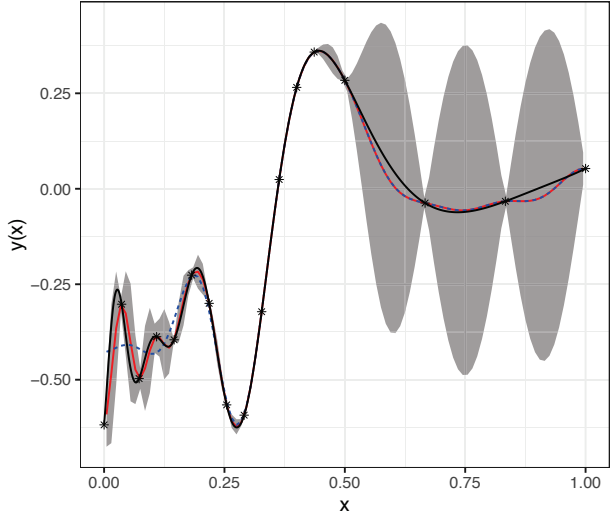}
  	\caption{\small \sl Predictions (in red) of the BJX test function $y(\bm{x})$  
	in \eqref{eq:bjx} and associated 95\% uncertainty intervals (as a 
	gray shadow)
			based on the CGP model.
			The dashed blue line is the estimate of the 
			global component $Y_{G}(\bm{x})$ under the
			CGP model.
		\label{fig:cgptgp}}
 \end{center}
\end{figure}

Figure~\ref{fig:cgptgp} plots CGP 
predictions of the BJX test function
\eqref{eq:bjx} based on the same $n = 17$ run
training data as above.   
For this example, the CGP predictions are clearly more accurate
across the input space than predictions under both kriging approaches shown 
in Figure~\ref{fig:bjxMPERK}.
The global predictor is smooth and captures the overall 
trend of the function well.
When the data are 
less volatile, as over the range $x \in [0.4,1]$,
the global predictor essentially interpolates the  data,
and the local predictions are approximately zero. 
Comparing uncertainty quantification between the methods,
for small $x$, CGP produces intervals that appear slightly wider than the
intervals under both kriging approaches. The CGP interval widths for large $x$
appear to fall in between the interval widths for kriging with constant and
cubic mean functions, and indicate a large amount of uncertainty about the
function in a region where it is essentially flat.

This paper introduces a Bayesian composite Gaussian process (BCGP)
model that modifies and extends the CGP model in several ways. 
The BCGP model extends the covariance structure used by
\citet{ba:12} to allow the global and local correlation functions to be differentially weighted. This 
provides the covariance function with greater flexibility to handle data sets where more or less
local adaptation is required.
An additional feature of the BCGP model is that it
introduces a new, flexible approach for handling the variance function
$\sigma^2(\bm{x})$.  Direct modeling of
the latent variance process is straightforward in the Bayesian context as it simply
requires a new level in a hierarchical model and an additional step in a
Markov chain Monte Carlo algorithm. We believe this direct approach to modeling will result 
in more accurate representations of uncertainty and will provide the model with additional flexibility 
for adapting to situations where the range of $y(\bm{x})$ varies significantly 
across the input space.  
More generally, by formulating the model in a 
Bayesian context we are able to
quantify uncertainty in the unknown model parameters.  By
fully 
integrating over the unknown parameters to predict $y(\bm{x})$,
the methodology allows one to 
fully quantify uncertainty in the predicted values.

After the BCGP model is introduced in Section~\ref{sec:bcgp},  Section~\ref{sec:comp} describes 
the computational algorithm we have developed for prediction and uncertainty quantification.
Section~\ref{sec:examples} performs prediction and uncertainty 
quantification for three examples. 
The first example is the BJX example, the second example is a $d=4$ setting that, visually, appears 
stationary, and the third example performs prediction for a  $d = 10$ 
analytic example of the wing weight of a light aircraft.


\section{The Bayesian Composite Gaussian Process Model} 
\label{sec:bcgp}

This section describes
a Bayesian composite Gaussian process (BCGP) model
that can be used to predict functions $y(\bm{x})$, $\bm{x} \in \mathcal{X}$, 
that, when viewed as a draw from a stochastic process, exhibit behavior 
consistent with non-stationarity. 
We assume that (perhaps after a suitable transformation)
the input space $\mathcal{X}$ is a $d$-dimensional, finite hyper-rectangle 
denoted by $[\bm{a}, \bm{b}]^d \equiv \prod_{j=1}^d[a_j, b_j]$, with 
$-\infty < a_j \leq x_j \leq b_j < +\infty $ for $j = 1, \ldots, d$.
As part of the
model specification below, we extend the GP notation for stationary processes, 
$\mbox{GP}(\beta_0,\sigma^2_Z,R(\cdot))$,
for use with nonstationary GPs by letting
$Y(\bm{x}) \sim \mbox{GP} (\mu(\bm{x}), C(\cdot, \cdot))$
indicate that $Y(\bm{x})$ follows a Gaussian process with
$E( Y(\bm{x}) ) = \mu(\bm{x})$ and covariance function
$C(\cdot, \cdot)$. Throughout, we assume that the training data have been
centered to have mean zero and scaled to have unit variance.

\subsection{Conditional Model}
\label{sec:likelihood}
The conditional (likelihood) component of the BCGP model assumes that 
$y(\bm{x})$ can be viewed as a realization from a random process $Y(\bm{x})$ 
that can be decomposed as
\begin{eqnarray}
	Y(\bm{x}) = Y_G(\bm{x}) + Y_L(\bm{x}) + \epsilon(\bm{x}), \;\;\;
		\bm{x} \in [\bm{a}, \bm{b}]^d,
		\label{eq:Ysum}
\end{eqnarray}
where $Y_G(\bm{x})$, $Y_L(\bm{x})$ and $\epsilon(\bm{x})$ are mutually
independent Gaussian processes.  As in the
CGP of \citet{ba:12}, the decomposition includes a global
component, $Y_G(\bm{x})$, and a local deviation
component, $Y_L(\bm{x})$. However, as seen below, our
model specification differs in significant ways. 
First, the model allows for the possible inclusion of a measurement error or 
nugget process $\epsilon(\bm{x})$
\citep[see][for a detailed discussion of the use of a nugget term in
GP  models for computer 
simulator output]{gram:12}. \citet{ba:12} argue that, due to the formulation of their CGP,
the local process may mimic a nugget term in some situations and hence do not
include such a term explicitly. We recognize that different practitioners will have 
different views on inclusion of a nugget component and note that, while we have
formulated the BCGP model to include the $\epsilon(\bm{x})$ for completeness, 
the nugget component can be easily removed if desired.

Conditional on model parameters 
$\bLambda = ( \beta_0,\omega,\boldsymbol\rho_G,\boldsymbol\rho_L,
\sigma^2_\epsilon, \sigma^2(\cdot) )$, we assume 
$Y_G(\bm{x}) \mid \bLambda \sim \mbox{GP}(\beta_0, C_G(\cdot, \cdot))$, 
where 
\begin{align}
	C_G(Y_G(\bm{x}_s),Y_G(\bm{x}_t)) = 
		\left\lbrace 
			\begin{array}{c l} 
				\sigma(\bm{x}_s)\sigma(\bm{x}_t)\, \omega \,
				G(\bm{x}_s-\bm{x}_t \mid \brho_G) 
				& , \, \bm{x}_s \neq  \bm{x}_t, \\ 
				\sigma^2(\bm{x}_s)\, \omega\, &, \, \bm{x}_s = \bm{x}_t,
			\end{array}
		\right. 
		\label{eq:globalCov}
\end{align}
$\sigma(\bm{x})$ is a positive function, $G$ is a global correlation function,
and $\omega \in [0, 1]$ is a weight. The local process is specified as 
$Y_L(\bm{x}) \mid  \bm{\Lambda} \sim GP(0, C_L(\cdot, \cdot))$, where
\begin{align}
	C_L(Y_L(\bm{x}_s),Y_L(\bm{x}_t)) = 
	\left\lbrace 
		\begin{array}{c l} 
			\sigma(\bm{x}_s) \sigma(\bm{x}_t)
			(1-\omega)L(\bm{x}_s - \bm{x}_t \mid \brho_L) & , \,
			\bm{x}_s\neq  \bm{x}_t, \\ 
			\sigma^2(\bm{x}_s)(1-\omega) &, \,
			\bm{x}_s = \bm{x}_t,
		\end{array}
	\right. 
	\label{eq:localCov}
\end{align}
and $L$ is a local correlation function. The process $\epsilon(\bm{x})$ is a 
mean zero Gaussian white noise process with variance $\sigma^2_\epsilon$.

The  functions $G$ and $L$ are
taken to be the
Gaussian correlation functions 
\begin{equation}
G\left( \bm{h} |\ \bm{\rho}_G \right) 
= 
\prod_{j=1}^{d} \rho_{G,j}^{K_G \left( h_j \right)^2}, 
\mbox{ and } 
L\left(\bm{h} |\bm{\rho}_{L} \right) 
= 
\prod_{j=1}^{d} \rho_{L,j}^{K_L \left( h_j \right)^2}\,  
\label{eq:gandl}
\end{equation}
with unknown parameters
$\brho_G = \l(\rho_{G,1},\ldots, \rho_{G,d} \r)$ and
$\brho_L = \l( \rho_{L,1},\ldots, \rho_{L,d} \r)$.
The quantities 
$K_G$ and $K_L$ are positive constants selected to 
enhance the numerical stability of operations on the
correlation matrices of the data; the values 
$K_G = K_L = 16$ are often appropriate when the data 
have been scaled to have unit variance.
 As with the CGP model, we take $Y_G(\bm{x})$ to be a smooth process that captures 
any global trend of $y(\bm{x})$, while $Y_L(\bm{x})$ adapts to local deviations.
The relative smoothness of draws from $Y_G(\bm{x})$ and $Y_L(\bm{x})$
is controlled by the global and local
correlation parameters $\brho_G$ and $\brho_L$.  
We force $Y_G(\bm{x})$ 
to be smoother than $Y_L(\bm{x})$ by embedding constraints 
in the joint prior distribution for $\brho_G$ and $\brho_L$.
 
Conditional on $\bm{\Lambda}$, 
the $Y(\bm{x})$ process (\ref{eq:Ysum})
can be equivalently specified as 
\begin{align}
	Y(\bm{x})\mid \bm{\Lambda} \sim 
	\mbox{GP}\left(\beta_0, C(\cdot,\cdot)\right), 
	\; \bm{x} \in [\bm{a},\bm{b}]^d,
	\label{eq:YGP}
\end{align}
 where 
 $\beta_0$ is the  overall mean, and 
\begin{align}
	C(Y(\bm{x}_s),Y(\bm{x}_t)) = 
	\left\lbrace 
		\begin{array}{c l} 
			\sigma(\bm{x}_s)\sigma(\bm{x}_t )
			 \left( \omega \, 
			G(\bm{x}_s-\bm{x}_t |\brho_G) 
			+ (1-\omega)\,
			L(\bm{x}_s-\bm{x}_t |\brho_L) 
			 \right) 
			& , \, \bm{x}_s \neq  \bm{x}_t,  \\ 
			\sigma^2(\bm{x}_s) + \sigma^2_\epsilon
			&, \, \bm{x}_s = \bm{x}_t\ . 
		\end{array}
	\right.
	\label{eq:covY}
\end{align}
The specification in (\ref{eq:Ysum})-(\ref{eq:localCov}) emphasizes 
the decomposition of the process into global, local and error components, 
while the specification in (\ref{eq:YGP})-(\ref{eq:covY}) emphasizes the roles of 
the parameters in the overall covariance function.  

As noted in Section~\ref{sec:likelihood}, the parameters $\brho_G$ and 
$\brho_L$ control the smoothnesses of the component processes
in $C(Y(\bm{\cdot}),Y(\bm{\cdot}))$.  The parameter  $\omega$
determines the extent that 
the model can make local adaptations to the global process;
no local adaption is allowed when $\omega = 1$. 
The final $C(Y(\bm{\cdot}),Y(\bm{\cdot}))$ parameter is $\sigma(\bm{x})$.
From (\ref{eq:covY}), $\mbox{Var}(Y(\bm{x}) \mid \bLambda) =
\sigma^2(\bm{x}) + \sigma^2_\epsilon$. In the applications we consider, 
$\sigma^2_\epsilon$ is typically small relative to the overall range of $y(\bm{x})$,
and hence $\sigma^2(\bm{x})$ plays the critical role in prediction and uncertainty
quantification with respect to the model variance. The conditional BCGP model
relies on knowing the form of 
$\sigma^2(\bm{x})$, which is typically not available in practice.
Rather than defining an algorithm for estimating 
$\sigma^2(\bm{x})$ as in \citet{ba:12}, 
we propose to model directly this function as an unknown random function by assuming
\begin{equation}
\label{eq:sigma_x_process}
	\log \sigma^2(\bm{x}) \mid \mu_V, \sigma^2_V, \brho_V \sim 
		\mbox{GP}(\mu_V, \sigma^2_V, G(\cdot \mid \brho_V)),
\end{equation}
where $G(\cdot \mid \brho_V)$ is the Gaussian correlation function 
in (\ref{eq:gandl}) with parameters $\brho_V$.

Modeling the variance function as a latent process provides a model-based approach
for flexibly estimating the volatility of the unknown function $y(\bm{x})$ 
across the input 
space. Specification of the model in this way introduces 
new, low-level parameters
$\mu_V$, $\sigma^2_V$ and $\brho_V$ that drive the unobserved process. 
Our model for the variance process is easily handled in our inferential and predictive
framework for two reasons. First, because we use 
MCMC methods
for inference and prediction, the fact that the variance process is simply a 
level in a 
Bayesian hierarchical model means that values of $\sigma^2(\bm{x})$ 
and of the hyper-parameters of this latent process can 
be updated by additional sampling steps. 
Second, due to the initial scaling of
the data, it is possible to use a 
prior distribution to center the 
parameters of the log Gaussian
process around reasonable values. 
This allows us
to anchor the $\sigma^2(\bm{x})$ function along a plausible trajectory 
while giving it 
the freedom
to adapt to information contained in the training data.

\subsection{Prior Model}
\label{sec:prior}
We complete the specification of the BCGP model with a prior distribution on
the unknown model parameters $\bm{\Lambda}$ that factorizes as follows:

\begin{equation}
		p(\beta_0) \ p(\omega)
		\ p(\sigma^2_\epsilon) \ p(\mu_V) \ p(\sigma^2_V) 
		\ \prod_{j=1}^d p(\rho_{L,j} \mid \rho_{G,j}) \ p(\rho_{G,j})
		\ p(\rho_{V,j}).
		\label{eq:prior}
\end{equation}
As is common in the literature, we assume a flat, location-invariant prior, 
$p(\beta_0) \propto 1$, for
the overall process mean. When the error process
is included in the model, we assign a gamma prior distribution to its variance,
$\sigma^2_\epsilon \sim \mbox{Gamma}(a_\epsilon, b_\epsilon)$, parameterized
so that $E(\sigma^2_\epsilon) = a_\epsilon b_\epsilon$. 
For data from a computer simulator 
we typically select
the hyperparameters so that $\sigma^2_\epsilon$ is, \emph{a priori}, 
close to zero with high probability(see Section~\ref{sec:examples} for examples).

The global correlation parameters are assumed to be independent of each other
with $\rho_{G,j} \sim \mbox{Beta}(\alpha_{G,j}, \beta_{G,j})$, for $j = 1, \ldots, d$.
While in principle one could chose different hyperparameters for each input
dimension, reflecting different \emph{a priori} beliefs about the function along
each input, in absence of such knowledge we typically set each $\alpha_{G,j}$
and $\beta_{G,j}$ equal to common
values $\alpha_G$ and $\beta_G$.
To enforce greater smoothness in the global process than in the local process in
each dimension, we specify the prior for the 
local correlation parameter  conditionally 
on the corresponding global parameter as a beta distribution
truncated to the interval $(0, \rho_{G,j})$:
\[
	\rho_{L,j} \mid \rho_{G,j} \stackrel{\mathrm{ind.}}{\sim}
		\mbox{TrBeta}(\alpha_{L,j}, \beta_{L,j}; 0, \rho_{G,j} ),
		\;\;\; j = 1, \ldots, d,
\]
The notation $X \sim \mbox{TrBeta}(\alpha, \beta; c, d)$ 
refers to a beta random variable truncated to the interval $(c,d)$
which has density
\[
	p(x) = \frac{\Gamma(\alpha + \beta)}
		{\Gamma(\alpha)\Gamma(\beta)}
		\frac{(x-c)^{\alpha-1}(d-x)^{\beta-1}}
		{(d - c)^{\alpha + \beta - 1}}, \;\;
		c \le x \le d,
\]
with mean $E(X) = c + \left( \frac{\alpha}{\alpha + \beta}  \right)(d-c)$
and variance
$Var(X) = \frac{\alpha \beta (d-c)^2}{(\alpha + \beta)^2(\alpha + \beta + 1)}$.
Lacking substantive prior information about the parameters
$\alpha_{L,j}$ and $\beta_{L,j}$ of $\rho_{L,j}$
we typically use common values $\alpha_L$ and $\beta_L$ across 
the $d$ inputs.

The prior for the parameter that weights the global and local correlation functions is 
taken to be 
$\omega \sim \mbox{TrBeta}(\alpha_\omega, \beta_\omega; L_\omega,
U_\omega)$, where $0 \leq L_\omega < U_\omega \leq 1$.
Often the prior for $\omega$ is truncated with $L_\omega = 0.5$ and
$U_\omega = 1$ to put more weight on the global process.

Finally, we assign a prior to the 
parameters $(\mu_V, \sigma^2_V, \brho_V )$ of the latent variable
process $\sigma^2(\bm{x})$ in (\ref{eq:sigma_x_process}) to have 
mutually independent components with marginals
\[
	\mu_V \sim \mbox{N}(\beta_V, \tau^2_V), \;\;\; 
	\sigma^2_V \sim \mbox{IG}(a_{\sigma^2_V}, b_{\sigma^2_V}), \;\;\;
	\rho_{V,j} \stackrel{\mathrm{iid}}{\sim} 
		\mbox{Beta}(\alpha_{\rho_{V,j}}, \beta_{\rho_{V,j}}), \;
		j = 1, \ldots, d,
\]
where $\mbox{IG}(a,b)$ represents the inverse gamma distribution with mean 
$(a-1)^{-1}b^{-1}$ when $a>1$.
To specify
values for the six hyper-parameters above 
recall that, ignoring the error variance $\sigma^2_\epsilon$, 
$\sigma^2(\bm{x})$ is the variance of the $Y(\bm{x})$ process.  
Assuming that the output $y(\bm{x})$ has 
been scaled to have zero sample mean and unit sample variance, we ``center''
our prior so that $\sigma^2(\bm{x}) \approx 1$ on average.  Setting
$\beta_V = -\frac{1}{10}$, $\tau^2_V = \frac{1}{10}$, $a_{\sigma^2_V} = 2+\sqrt{\frac{1}{10}}$, 
and $b_{\sigma^2_V} =  \frac{100}{1+\sqrt{\frac{1}{10}}}$ encourages the $\sigma^2(\bm{x})$
process to stay near unity 
on average while allowing the data to suggest regions of the input
space where $\sigma^2(\bm{x})$ should be larger or smaller.
Lastly, the hyperparameters $\alpha_{\rho_{V,j}}$ and 
$\beta_{\rho_{V,j}}$ can be chosen to control the smoothness of the latent variance process.
In general, we expect this process to be fairly smooth, which suggests
picking values that encourage
high correlation. If there is a strong prior belief that the 
unknown $y(\bm{x})$ may be 
best modeled as a stationary process, then setting the $\alpha_{\rho_{V_j}}$ and 
the $\beta_{\rho_{V_j}}$ to ensure that the $\rho_{V_j}$ are close to 1, 
will encourage 
a (nearly) constant variance function.
Setting $\alpha_{\rho_{V_j}} = \beta_{\rho_{V_j}} = 1$ gives a non-informative 
$\mbox{Unif}(0,1)$ distribution.

\section{Computational Algorithms for Inference and Prediction} 
\label{sec:comp}

This section describes the computational algorithms we have developed
for inference and prediction under the BCGP model. 
Assume that the 
unknown function $y(\bm{x})$ has been sampled at $n$ training data 
sites in the input
space, denoted $\bx_i$, $i = 1, \ldots, n$, and 
$\by = (y(\bx_1), \ldots, y(\bx_n))^\top$ is the associated
vector of computed values of $y(\bm{x})$.
To simplify notation, let 
$\bV = (\sigma^2(\bx_1), \ldots, \sigma^2(\bx_n))^\top$ 
be the random vector of unknown values of the 
variance process at the training data
locations.  We augment the collection of parameters 
$\bLambda$ introduced in
Section~\ref{sec:likelihood} to include all unknown quantities so that now
$\bLambda = (\beta_0$, $\omega$, $\brho_G$,  $\brho_L$, 
$\sigma^2_\epsilon$, $\bV$, $\mu_V$, $\brho_V$, $\sigma^2_V)$.
The posterior distribution of all unknown quantities $\bLambda$ has density function 
$p(\bLambda \mid \by) \propto p(\by \mid \bLambda)p(\bLambda)$, where 
$p(\bLambda)$ is
the prior specified in Section~\ref{sec:prior}.  The likelihood  
$p( \by \mid \bLambda)$ is derived from the conditional model specified in 
\eqref{eq:YGP} and \eqref{eq:covY}, which implies that
$
	\by \mid \bLambda  \sim  \mbox{N} \left( \beta_0 \bone, \bC \right),
$
where the $n \times n$ covariance matrix $\bC$ has $(i,j)^{th}$
element, $1 \leq  i, j \leq n$, 
\begin{equation}
	C_{ij} =  \sigma \left(  \bx_i \right)
		\sigma \left( \bx_j  \right) 
		 \left(\omega G\left(\bx_i - \bx_j \mid \brho_G \right) 
			+ (1-\omega) L  \left(\bx_i - \bx_j \mid \brho_L \right) \right) 
			+ \delta_{i,j} \sigma^2_\epsilon, \label{eq:Ccovmatrix}
\end{equation} 
and $\delta_{\cdot,\cdot}$ is the Kronecker delta function.

The posterior density $p(\bLambda \mid \by)$ is 
difficult to compute in closed form.  For inferential and predictive purposes, we obtain samples 
from the posterior via a Markov chain Monte Carlo (MCMC) algorithm.
We update parameters and the $\bV$ values
based on their full conditional distributions
either by Gibbs updates---sampling directly from full conditional distributions---or by 
Metropolis--Hastings updates---sampling from proposal distributions and accepting or 
rejecting the proposed draws based on full conditional distributions.  Some updates are relatively
straightforward, while others---in particular, the update of the latent variance process 
$\bV$---require special attention 
in order to ensure good mixing of the chain.

The MCMC algorithm starts at an initial value $\bLambda^{[0]}$ and, at each iteration $t$, 
the elements of $\bLambda$ are updated according to the 9 steps below.
The chain can be initialized with any $\bLambda^{[0]}$ satisfying 
$p(\bLambda^{[0]} \mid \by ) > 0$.
For simplicity the following notation is used.
At any step in the sampler during iteration $t$, 
the notation $\bLambda^{[t]}$  
represents a vector containing the newly-sampled values 
for any parameters that have already 
been updated in the (partial)
sweep through the steps at iteration $t$.  Similarly, for steps with Metropolis--Hastings  
updates, the notation $\bLambda^\prime$ should be understood to mean 
$\bLambda^{[t]}$ where 
the parameters currently being updated are 
replaced with proposed values.
For a generic parameter $\theta$, the notation
$\bLambda_{-\theta}^{[t]}$ should be understood to be the up-to-date version of the parameter vector
without component $\theta$.  Unless otherwise specified, for all  
proposal distributions used in 
 Metropolis--Hastings updates, if the proposed value is 
 $\theta^\prime$, the updated value is 
taken to be
\begin{align}
	\theta^{[t+1]} = \left\lbrace \begin{array}{c l} \theta' & \mbox{with probability } \min\left\lbrace 1, 
			\frac{ p(\bm{\Lambda}' \mid \by )}{ p(\bm{\Lambda}^{[t]} \mid \by ) } \right\rbrace, \\ 
	\theta^{[t]} &\mbox{with probability } 1 - \min\left\lbrace 1, 
			\frac{ p( \bm{\Lambda}' \mid \by )}{ p(\bm{\Lambda}^{[t]} \mid \by )} \right\rbrace. 
			\end{array}\right.
	\label{eq:MHrat}
\end{align}
In our MCMC algorithm, a Metropolis--Hastings 
update for a parameter $\theta$ relies on a 
\emph{calibrated proposal width} $\Delta_\theta$ to help ensure reasonable mixing of the chain. 
Section~\ref{sec:calprop} provides details of the calibration scheme.

At iteration $t$ in the MCMC algorithm, the parameters are updated according to the following steps.
\begin{description}
\item[Step~1:] Update $\beta_0$ by sampling $\beta_0^{[t+1]}$ directly from its full conditional distribution,
\[
	\beta_0 \mid \bLambda^{[t]}_{-\beta_0}, \by \sim 
		\mbox{N}\left( \left(\bone^\top\bC^{[t] -1} \bone \right)^{-1} \bone^\top\bC^{[t] -1}\by,
			\left(\bone^\top\bC^{[t] -1}\bone \right)^{-1}  \right),
\]
where $\bone$ is a vector of ones and $\bC^{[t]}$ is the covariance matrix with elements (\ref{eq:Ccovmatrix})
evaluated at the training data points $\bx_i$ using the parameters $\bLambda^{[t]}_{-\beta_0}$.

\item[Step~2:] Update $\omega$ by proposing $\omega^\prime$ from a 
$\mbox{Unif}\left(\omega^{[t]}-\Delta_\omega, \omega^{[t]}+\Delta_\omega \right)$ distribution and 
using (\ref{eq:MHrat}) to determine the value of $\omega^{[t+1]}$.

\item[Step~3:] Update the global correlation parameters $\rho_{G,1}, \ldots, \rho_{G,d}$ one-at-a-time 
(conditioning on the others) by proposing 
$\rho_{G,j}^\prime$ from a $\mbox{Unif}\left(\rho_{G,j}^{[t]} - \Delta_{\rho_{G,j}}, \rho_{G , j}^{[t]}
+\Delta_{\rho_{G, j}} \right)$ 
distribution and using (\ref{eq:MHrat}) to determine the value of each $\rho_{G, j}^{[t+1]}$.

\item[Step~4:] Update the local correlation parameters $\rho_{L, 1}, \ldots, \rho_{L, d}$ one-at-a-time 
(conditioning on the others) by proposing
$\rho_{L, j}^\prime$ from a $\mbox{Unif}\left(\rho_{L, j}^{[t]} - \Delta_{\rho_{L, j}}, \rho_{L, j}^{[t]} + 
\Delta_{\rho_{L, j}} \right)$ distribution and using (\ref{eq:MHrat}) to determine the value of each 
$\rho_{L, j}^{[t+1]}$.

\item[Step~5:] Update $\sigma^2_\epsilon$ by proposing ${\sigma^{2}_\epsilon}^\prime$ from a 
$\mbox{Unif}\left(\sigma^{2^{[t]}}_\epsilon - \Delta_{\sigma^2_\epsilon}, \sigma^{2^{[t]}}_\epsilon + 
\Delta_{\sigma^2_\epsilon} \right)$ distribution and using (\ref{eq:MHrat}) to 
determine the value of ${\sigma^2}^{[t+1]}_\epsilon$.

\item[Step~6:] Update $\mu_V$ by sampling $\mu_V^{[t+1]}$ directly from its full conditional distribution,
\[
	\mu_V \mid \bLambda^{[t]}_{-\mu_V}, \by \sim \mbox{N}\left(m,v\right),
\]
where $v^{-1} = 1/\tau^2_V + \bone^\top {\bR_t^{[t]}}^{-1}\bone /{\sigma^{2}_V}^{[t]}$ and
$m = v (\beta_V/\tau^2_V + \bone^\top {\bR_t^{[t]}}^{-1}\bW^{[t]} /{\sigma^{2}_V}^{[t]})$,
$\bW^{[t]} = \log \bV^{[t]} = \left( \log \sigma^{2^{[t]}}(\bx_1),\ldots, \log \sigma^{2^{[t]}}(\bx_n) \right)^\top$, and 
$\bR_t^{[t]}$ is the correlation matrix for the $\log\;\sigma^2\left(\bm{x}\right)$ process evaluated at the training data locations
with elements
\[
	\bR_{t_{ij}}^{[t]} = G\left( \bx_i - \bx_j  \mid \brho_V^{[t]}\right).
\]

\item[Step~7:] Update $\sigma^2_V$ by sampling $\sigma^{2^{[t+1]}}_V$ directly from its full conditional distribution,
\[
	\sigma^2_V \mid \bLambda^{[t]}_{-\sigma^2_V}, \by \sim IG\left( \frac{n}{2} + a_{\sigma^2_K}, 
		\left( \frac{1}{2}\left( \bW^{[t]} - \mu_V^{[t+1]}  \right)^\top \left(\bR^{[t]}_t\right)^{-1} 
			\left( \bW^{[t]} - \mu_V^{[t+1]}  \right) + \frac{1}{b_{\sigma^2_K}} \right)^{-1}  \right).
\]

\item[Step~8:] Update $\rho_{V, 1}, \ldots, \rho_{V, d}$ one-at-a-time (conditioning on the others)
by proposing $\rho_{V, i}^\prime$ from a $\mbox{Unif}\left(\rho_{V, i}^{[t]} - 
\Delta_{\rho_{V, i}}, \rho_{V, i}^{[t]} + \Delta_{\rho_{V, i}}\right)$ distribution
and using (\ref{eq:MHrat}) to determine 
the value of each $\rho_{V, i}^{[t+1]}$.

\item[Step~9:] Update $\bV = (\sigma^2(\bx_1), \ldots, \sigma^2(\bx_n))^\top$
as described in Section~\ref{sec:updateW}.

\end{description}

In practice, the MCMC algorithm is run for 
three sets of iterations.  The first set   
are {\em calibration} iterations in
which a fixed number of iterations are made 
in which the proposal widths 
$\Delta_\theta$ are determined for the subsequent runs
 (see Section~\ref{sec:calprop}).  
After calibration, the chain is run for an 
additional {\em burn-in} period.  The final set of
iterations are  
$n_{mcmc}$ {\em production} iterations that produce 
samples $\bLambda^{[1]}, \ldots, \bLambda^{[n_{mcmc}]}$ 
from the posterior distribution $p(\bLambda \mid \by)$. The 
samples can be used for predictive inference as described in Section~\ref{sec:pred}.  

\subsection{Updating the latent variance process $\bV$}
\label{sec:updateW}
Updating the latent variance process at the training data locations 
$\bx_1, \ldots, \bx_n$ requires special attention.  The full conditional posterior 
distribution of neither $\bV$ nor its logarithm, $\bW$, are standard distributions and so sampling updated
values directly is difficult.  When the number of training 
data locations, $n$, is not ``too large'', 
we can use a Metropolis--Hastings update for the full vector $\bV$ 
by sampling from a proposal process at the training data locations and accepting the proposed
move with the appropriate probability.  
While straightforward in principle, the proposal process must be 
constructed carefully in order to ensure acceptance rates that result in appropriate mixing of the chain.   
When $n$ is large it is difficult to accept the entire vector of proposed values $\bV^\prime$
unless the proposal vector is 
very close to the current vector, which inhibits mixing. 

With this in mind, we describe two different methods for updating the 
latent variance process.  The first method is designed to work well when $n$ is ``small'', while the second method 
is constructed to produce reasonable mixing when $n$ is large.
When the number of training data locations is small, say $n < 20$, we recommend updating $\bV$ by sampling
\begin{equation}
	\bW^\prime \sim N\left( \bW^{[t]}, \bK^{[t]}_W\right), \label{eq:smallNupdate}
\end{equation}
where $\bW^{[t]} = \log \bV^{[t]} = \left( \log \sigma^{2^{[t]}}(\bx_1),\ldots, \log \sigma^{2^{[t]}}(\bx_n) \right)^\top$,
$\bK^{[t]}_{W_{ij}} = \tau^2 G\left( \bx_i - \bx_j \mid \brho^{[t+1]}_V\right)$, and $\tau^2$ is a predefined value 
that controls the variance of the proposal distribution. The 
proposal distribution (\ref{eq:smallNupdate}) 
is centered around the current 
value at each training data location. The variance parameter $\tau^2$ 
should be chosen so that the proposed 
values, $\bW^\prime$, are (1) similar enough to the current values, 
$\bW^{[t]}$, 
to have a useful acceptance rate while (2)
still allowing the $\bW^\prime$ to be sufficiently 
different from the $\bW^{[t]}$ 
values so the support of the posterior distribution can 
be fully explored.  After appropriately accounting for the transformation, the acceptance probability for the proposed 
value $\bV^\prime = (e^{W^\prime_1}, \ldots, e^{W^\prime_n})^\top $ is $\min\left\lbrace 1, p  \right\rbrace$, where 
\begin{equation}
	p =  \frac{ p(\by \mid \bLambda' )}{p(\by \mid \bLambda^{[t]})} \times 
		\frac{exp\left( -\frac{1}{2}\left( \bW' - \mu_V^{[t+1]}\bone \right)^\top {\bK^{[t]}}^{-1}
		\left( \bW' - \mu_V^{[t+1]}\bone \right) \right) }{exp\left( -\frac{1}{2}
		\left( \bW^{[t]} - \mu_V^{[t+1]}\bone \right)^\top {\bK^{[t]}}^{-1}\left( \bW^{[t]} - \mu_V^{[t+1]}\bone \right) \right)}.
	\label{eq:forVPred}
\end{equation}

When the number of training data locations is large, say $n \geq 20$, we recommend an alternate approach to 
updating $\bV$.  Rather than updating all $n$ elements of $\bV$ together, we instead randomly select a focal point 
from the design space and then update the variance process at a cluster of $n_{prop}$ training locations closest to 
the chosen focal point conditionally on the current value of the variance process at all other training data locations. 
After updating the process at that cluster of points, we randomly select another focal point in the design space and 
repeat the process.  At each iteration in the overall MCMC algorithm, the process is repeated so that $m$ total 
focal points are sampled, and the final vector $\bV$ after the $m$ cycles is retained as new state $\bV^{[t+1]}$ 
in the Markov chain.  While $m=1$ yields a valid MCMC algorithm, we expect setting $m>1$ should improve mixing 
of the chain. The following steps describe the details of this process.
\begin{description}
\item[Step~9a:] Select a focal point 
uniformly at random from the
$d$-dimensional, hyper-rectangular input space $[\bm{a}, \bm{b}]^d$.

\item[Step~9b:] Select the $n_{prop}$ training data locations closest to the randomly selected focal point. In practice, 
we have found that choosing $n_{prop} = 15$ works well. Denote these points by $\mathbf{\bar{x}}$ and the remaining
training data points by ${\mathbf{\ubar{x}}}$. 

\item[Step~9c:] Propose new values $\bW^\prime(\mathbf{\bar{x}})$ by sampling from the distribution obtained by conditioning (\ref{eq:smallNupdate})
on the current values $\bW^{[t]}(\mathbf{\ubar{x}})$:
\[
	\bW'\left(\mathbf{\bar{x}}\right) \mid \bW^{[t]} \left(\mathbf{\ubar{x}}\right), \brho_V^{[t+1]} \sim 
		\mbox{N}\left( \bW^{[t]} \left(\mathbf{\bar{x}}\right), \tau^2\left( \bR_{\overline{W}} -  \bR_{\overline{W},\underline{W}}^\top 
		\bR_{\underline{W}}^{-1} \bR_{\overline{W},\underline{W}}\right)  \right),
\]
where $\bR_{\overline{W}_{ij}} = G\left( \mathbf{\bar{x}}_i - \mathbf{\bar{x}}_j \mid \brho_V^{[t+1]} \right)$ is the 
$n_{prop} \times n_{prop}$ correlation matrix for the log GP between the proposal locations, 
$\bR_{\underline{W}_{ij}} = G\left( \mathbf{\ubar{x}}_i - \mathbf{\ubar{x}}_j \mid \brho_V^{[t+1]} \right)$ 
is the $\left(n- n_{prop}\right) \times \left(n-n_{prop}\right)$ correlation matrix for the log GP between the locations 
in $\mathbf{\ubar{x}}$, and 
$\bR_{\overline{W}, \underline{W}_{ij}} = G\left( \mathbf{\ubar{x}}_i - \mathbf{\bar{x}}_j \mid \brho_V^{[t+1]} \right)$ 
is an $\left(n-n_{prop}\right) \times n_{prop}$ matrix with each column containing the correlation for the log GP 
between a proposal point and each of the locations in $\mathbf{\ubar{x}}$.

\item[Step~9d]: Update the elements of $\bW^{[t]}$ corresponding to the locations $\mathbf{\bar{x}}$ with
the values $\bW^\prime(\mathbf{\bar{x}})$ with probability $\min\{1,p\}$, where $p$ is as in 
\eqref{eq:forVPred}; otherwise, do not change $\bW^{[t]}$.

\item[Step~9e:] After repeating Steps~(9a)-(9d) $m$ times, set
$\bW^{[t+1]} = \bW^{[t]}$ and $ \bV^{[t+1]} = \exp\left(\bW^{[t+1]}\right) = 
\left( \sigma^{2^{[t+1]}}\left(\bx_1\right),\ldots,\sigma^{2^{[t+1]}}\left(\bx_n\right) \right)^\top.$
\end{description}
In our examples, we typically set $m$ so that $m \times n_{prop} > n$, which has resulted in satisfactory 
mixing of the chain.

\subsection{Calibrating the proposal widths}\label{sec:calprop}
The Metropolis--Hastings updates described above rely on proposal 
widths $\Delta_\omega$, 
$\Delta_{\rho_{G, j}}$, $\Delta_{\rho_{L, j}}$, $\Delta_{\sigma^2_{\epsilon}}$ and $\Delta_{\rho_{V, j}}$. 
Appropriate values must be chosen in order to ensure good mixing of the chain.  We use an automated method 
to calibrate these proposal widths with the goal of selecting widths that result in acceptance rates of between approximately $0.2$ and $0.4$.  It has been shown theoretically that in specific contexts acceptance rates in this 
range lead to chains with good convergence and mixing properties \citep[e.g.,][]{gelm:96, robe:97, robe:01}; 
empirical evidence in many different model and data settings suggests these rates are generally desirable.

To adaptively calibrate the proposal widths, we initially run the MCMC algorithm with user-specified widths 
$\Delta_\theta$. The proposal widths can be different for each parameter $w$, $\rho_{G, j}$, \emph{etc}.  
After $n_{adapt}$ iterations, we compute the empirical acceptance rates separately for each parameter with a 
proposal width and compare these acceptance rates to a range of target rates (our implementations uses the 
range $[0.25, 0.40]$).  If any individual empirical rate $acceptRate$ is outside of this range the proposal 
width for that parameter is updated to be $\Delta_\theta := \Delta_\theta * acceptRate / c$, where $c$ is a 
specific target rate.  Under this scheme, a proposal width will be increased when the empirical acceptance rate is 
too high and decreased when too low relative to the target. After updating the $\Delta_\theta$, the MCMC algorithm 
is continued for another $n_{adapt}$ iterations. The adaptation scheme is terminated after a total of $numUpdates$
adaptation periods.  Section~\ref{sec:examples} provides examples of how we have implemented this approach in
practice.

Because the transition kernel is potentially changing throughout the adaptation period, we discard all samples at 
the end of the $numUpdates$ adaptation periods and start a new MCMC run using the final state of the chain as 
the starting values $\bLambda^{[0]}$ and fixing the calibration widths $\Delta_\theta$ at their final values.  As we do not assume
that we start 
the chain from stationarity, we typically allow for an additional burn-in period before collecting production
samples from the posterior.

\subsection{Prediction and Uncertainty Quantification}\label{sec:pred}

A primary objective is to use the methodology to predict the output of a computer simulator (or other
source) at new input values. Quantification of uncertainty about these predictions is also desired. Focusing on a 
particular (single) input location $\bx_*$, predictive inference under the BCGP model is obtained via the 
posterior predictive distribution
\[
	p(y(\bx_*) \mid \by) = \int p(y(\bx_*) \mid \by, \bLambda)\ p(\bLambda \mid \by) \ d \bLambda,
\]
where the unknown parameters are integrated 
over their ``likely'' values as 
specified by the posterior distribution.  The point
prediction is taken to be the posterior predictive mean $E(Y(\bx_*) \mid \by)$.  Uncertainty about the unknown 
value of $y(\bx_*)$ 
is quantified by a $(1-\alpha)\times 100\%$ posterior predictive interval 
computed as lower and upper 
$\alpha/2$ percentiles of the posterior predictive distribution.

To compute the predictions, note that
the conditional distribution of $Y(\bx_*)$ given $\bLambda$ and $\by$ is
\begin{equation}
	Y(\bx_*) \mid \bLambda, \by \sim N\left( \beta_0 + \bC_*^\top \bC^{-1} (\by - \beta_0 \bone) \; , \;
		 \sigma^2\left( \bx_*\right) + \sigma^2_\epsilon - \bC_*^\top \bC^{-1} \bC_* \right),
		\label{eq:condpred}
\end{equation}
where $\bC$ is the covariance matrix at the training data locations with elements calculated as in 
(\ref{eq:Ccovmatrix}) and  
$\bC_* = (C_{*1}, \ldots, C_{*n})^\top$ is the vector of covariances
between the process at the prediction input $\bx_*$ 
and the process at the training input locations; these elements 
are 
\[
	C_{*i} = \sigma(\bx_*)\sigma(\bx_i) (w G(\bx_* - \bx_i \mid \brho_G) + 
								(1-w) L(\bx_* - \bx_i \mid \brho_L)),
\]
for $i = 1,\ldots,n$. The conditional distribution 
(\ref{eq:condpred}) can be used to construct the 
Rao--Blackwellized Monte Carlo 
estimate 
\begin{equation}
	\widehat{E}(Y(\bx_*) \mid \by) = \frac{1}{n_{mcmc}} \sum_{t=1}^{n_{mcmc}}
		\left( \beta_0^{[t]} + \bC_*^{[t]^{\top}} \bC^{[t]^{-1}} (\by - \beta_0^{[t]}\bone)
		\right)
		\label{eq:RBpred}
\end{equation}
of $E(Y(\bx_*) \mid \by)$ using the posterior samples obtained with 
the MCMC algorithm, 
where quantities superscripted by $[t]$ are computed using the $t$-th draw 
of the parameters, $\bLambda^{[t]}$.

Computing $\bC_*^{[t]}$ requires the term 
$\sigma^{{[t]}}(\bx_*)$, the square root of the  $t^{th}$
\emph{a posteriori} sample of the 
latent variance function at 
$\bx_*$. While the method for updating the latent variance process described 
in Section~\ref{sec:updateW} produces samples of the latent variance process at the training data locations, it 
does not automatically produce samples at the prediction location.  If the prediction location is known in advance, 
the approach described in Section~\ref{sec:updateW} can be modified 
to include the prediction location together 
with the training data locations.
The resulting $\sigma^{[t]}(\bx_*)$ can be saved for prediction. 

If  the prediction location is not known before the MCMC algorithm is 
run, samples can be obtained
after the end of the MCMC run
by simulating from the appropriate conditional distribution. 
In more detail, letting 
$W_* = \log \sigma^2(\bx_*)$, we have
\begin{equation}
	W_* \mid \bLambda, \by \sim N \left( \mu_V + \bR_*^\top \bR^{-1}(\bW_t - \mu_V \bone) \; , \;
		\sigma_V^2 (1 - \bR_*^\top \bR^{-1} \bR_*) \right),
		\label{eq:Wcond}
\end{equation}
where $\bR$ is the $n  \times n$ correlation matrix 
of the latent log GP evaluated at the training data locations
and has $ij$-th element $\bR_{ij} = G(\bx_i - \bx_j \mid \brho_V)$.
The term 
$\bR_*$ is the $n \times 1$ vector
of covariances between the latent 
log GP at $\bx_*$ and each of the training data inputs
$\bx_i$, i.e., 
$\bR_{*i} = G(\bx_* - \bx_i \mid \brho)$, for $i= 1,\ldots n$.  
Each posterior sample $\bLambda^{[t]}$, $t= 1,\ldots, n_{mcmc}$, 
is used to generate a draw $W^{[t]}_*$ 
from (\ref{eq:Wcond}); transforming $W^{[t]}_*$ yields 
$\sigma^{[t]}(\bx_*)$ which is required to evaluate 
the vector $\bC_*^{[t]}$ in (\ref{eq:RBpred}).

Uncertainty about the unknown value of the function at the 
input $\bx_*$ is quantified via a 
$(1-\alpha) \times 100\%$ posterior predictive interval computed by finding 
the upper and lower $\alpha/2$ percentiles 
of the posterior predictive distribution. As discussed in \citet{davi:15}, 
Rao--Blackwellized Monte Carlo estimates of 
these quantities are more difficult to compute than the Rao--Blackwellized 
point predictions described above. 
A computationally simpler approach is to obtain the percentiles 
required to construct the predictive intervals 
by first obtaining samples $Y^{[t]}(\bx_*)$,
$t = 1, \ldots, n_{mcmc}$, 
from the conditional predictive distribution 
(\ref{eq:condpred}) using the $\bLambda^{[t]}$ and
$\sigma^{[t]}(\bx_*)$ samples. 
Then, for example, the $2.5^{th}$ and $97.5^{th}$ percentiles 
of this set of samples are 
Monte Carlo estimates of the endpoints of the $95\%$
posterior predictive interval. Note that averaging the samples 
$Y^{[t]}(\bx_*)$ together would also produce a valid estimate of $E(Y(\bx_*) \mid \by)$; however, when 
computationally feasible, we prefer the less-noisy, Rao--Blackwellized approach.

The methods of prediction and uncertainty quantification described above are specific to a single new input 
location $\bx_*$. Point-wise prediction and uncertainty quantification at several new input locations $\bx_{*k}$, 
$k = 1, \ldots, n_p$, is easily achieved by implementing the methods separately at each location.

\subsubsection{Global and Local Components of Prediction}

\citet{ba:12} emphasized that
predictions under the CGP model 
can be decomposed into \emph{global} and
\emph{local} components.
Decomposing predictions in this way allows one to visually assess how the behavior of the 
unknown function changes over the input space, e.g.~by finding regions where large local adaptations are 
necessary. 

Posterior predictions under the BCGP model can be similarly
decomposed by rewriting the 
conditional posterior predictive mean (\ref{eq:condpred}) as
\begin{align}
	E(Y(\bx_*) \mid \bLambda, \by) &=  \beta_0 + \bC_*^\top \bC^{-1} (\by - \beta_0 \bone) \nonumber \\
		&= \beta_0 + \bC_{G_*}^\top \bC^{-1} (\by - \beta_0 \bone) + 
		\bC_{L_*}^\top \bC^{-1} (\by - \beta_0 \bone) + \bC_{\epsilon_*}^\top \bC^{-1}
		(\by - \beta_0 \bone). 
\label{eq:preddecomp}
\end{align}
The representation in (\ref{eq:preddecomp}) is due to the fact that the process covariance between the training
data locations and the prediction location can be decomposed into global, local and error components.
The elements of the vector $\bC_{G_*}$ are computed using the global covariance function (\ref{eq:globalCov})
and represent the global component of the covariance between $Y(\bx_*)$ and $Y(\bx_i)$, $i = 1, \ldots, n$;
$\bC_{L_*}$ is defined similarly for the local component of the covariance function \eqref{eq:localCov}.
The vector $\bC_{\epsilon_*}$ corresponds to the ``error'' component of the model. All elements
of this vector will be zero unless we are predicting at one of the training data locations, i.e.~$\bx_* = \bx_k$ for some 
$k \in \{1, \ldots, n\}$, in which case the $k$th element of the vector will be $\sigma^2_\epsilon$.

Using this decomposition, the BCGP predictor (\ref{eq:RBpred}) that averages over the posterior distribution of the 
unknown parameters can be re-written as
\[
	\widehat{E}(Y(\bx_*) \mid \by) = \frac{1}{n_{mcmc}} \sum_{t=1}^{n_{mcmc}}
		\left( \widehat{y}_G^{[t]}(\bx_*) + \widehat{y}_L^{[t]}(\bx_*) + 
		\widehat{y}_{\epsilon}^{[t]}(\bx_*) \right),
\]
where $\widehat{y}_g(\bx_*) = \beta_0 +  \bC_{G_*}^\top \bC^{-1} (\by - \beta_0 \bone)$,
$\widehat{y}_L(\bx_*) = \bC_{L_*}^\top \bC^{-1} (\by - \beta_0 \bone)$, and
$\widehat{y}_{\epsilon}(\bx_*) = \bC_{\epsilon_*}^\top \bC^{-1} (\by - \beta_0 \bone)$
can be viewed as the global, local and error components of the overall prediction, respectively.
The error component $\widehat{y}_{\epsilon}(\bx_*)$ will be zero except when making a prediction at 
one of the training data locations.

\section{Examples} 
\label{sec:examples}

This section applies BCGP prediction to three examples.  The first is the
BJX function introduced in Section~\ref{sec:intro}.   The second is
a $d=4$ example using the output of a heat exchange simulator code.  
The final example
uses output from a $d = 10$ analytic function for the wing weight
of a light aircraft.

\noindent
{\bf Example~4.1} 
Consider BCGP prediction of the $d =1$  non-stationary 
 $y(x)$ of \citet{ba:12} and 
\citet{xion:07}
given by equation 
\eqref{eq:bjx}. 
Prediction and uncertainty quantification of $y(x)$
are based on the BCGP model with the  
prior form described in Section~\ref{sec:prior}
and the following hyperparameter specifications. 
BCGP  was run using 
$60,000 = 60 \times 1000$ iterations for calibration,
followed by 4,000 burn-in iterations, and 5,000 production
iterations. The $\omega$ prior was taken to
be the  $Beta(4, 6)$ distribution truncated to $[0.5, 1.0]$
which yields an $\omega$ prior mean of $0.7$ 
and prior standard deviation of 0.074. The prior for  
$\rho_{G,1} = \rho_{G}$ was Beta(1.0,0.4).
The conditional distribution of $\rho_{L,1}=\rho_{L}$ given
$\rho_{G}$ was taken to be 
Beta(1.0,1.0) truncated to $[0, \rho_{G}]$.  
Thus the prior mean for $\rho_{G}$
is $1/1.4 = 0.71$ while 
the conditional prior mean for $\rho_{L}$ 
is $0.5 \times \rho_{G}$.

\begin{figure}[t] 
\centering 
 \includegraphics[width=.55\linewidth]
{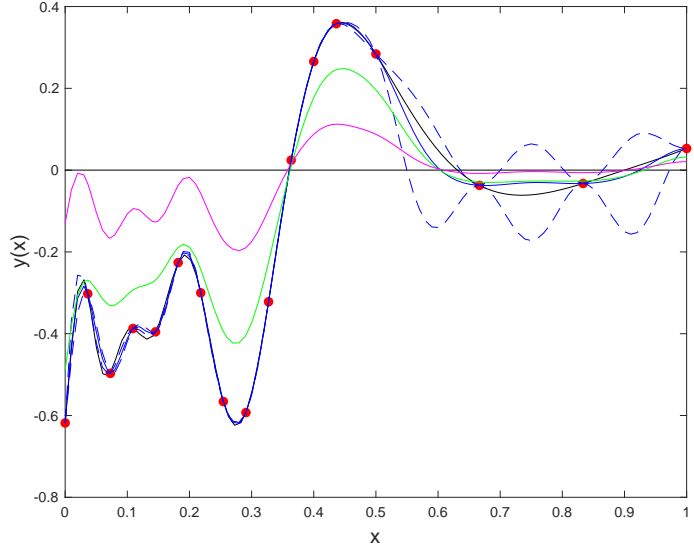} 
\caption{
\small \sl
Prediction and  95\% uncertainty bounds
for the BJX function, $y(x)$, in Example~4.1 (solid black):
the BCGP predictor of $y(x)$ (solid blue); 
95\% UQ limits of $y(x)$ (dashed blue);
estimated posterior mean
of the ${Y}_G(x)$ process (solid green);
estimated posterior mean of the
of the ${Y}_L(x)$ process (solid magenta).
}
\label{fig:Preds_for_BJX}  
\end{figure}

The BCGP predictor and associated 95\% point-wise 
uncertainty bounds are shown in  
Figure~\ref{fig:Preds_for_BJX}. 
As desired and seen by visual inspection,
the bounds for $x > 0.5$
are not greatly influenced by the relatively large variations in
$y(x)$ for $x < 0.5$.  In contrast, the 95\% bands
produced by kriging predictors having constant mean or cubic mean
(Figure~\ref{fig:bjxMPERK})
contain a substantial amount of uncertainty when $x > 0.5$
that is caused by the stationarity assumption 
coupled by the training data with $x <0.5$.
Figure~\ref{fig:Preds_for_BJX} also shows the posterior
means of the 
predicted global and local deviations (and their sum, the 
predicted BJX function).  The solid green global mean estimate,
$\widehat{y}_G(x)$, 
captures  the global trend well and is not myopically 
concerned with the small variations in the function
where $x <0.5$.  The solid purple local
deviation estimate, $\widehat{y}_L(x)$, is centered about zero
and fluctuates more rapidly in regions were more adaption is required.
The sum of the two components
produces the underlying $y(x)$.  

\begin{table}[b]
\begin{center}
{\singlespacing
\begin{tabular}{|c|c|}
\hline
\multicolumn{1}{|c|}{Predictor} &
\multicolumn{1}{c|}{RMSPE} \\
\hline  \hline 
Kriging/Constant \ mean & 0.067 \\
Kriging/Cubic \ mean &  0.061 \\
CGP & 0.023 \\
BCGP & 0.014 \\ \hline
\end{tabular}
\caption{\small \sl Root Mean Squared Prediction Errors (RMSPEs) over 
the grid 0.0(0.01)1.0 for four predictors of the
BJX function.}
\label{ta:rmspe}
}
\end{center}
\end{table}

Table~\ref{ta:rmspe}
compares the prediction accuracy of the BCGP predictor
with those of the kriging and CGP predictors.
While the BCGP predictor has the smallest root 
mean squared prediction error (RMSPE), the accuracy
of the CGP predictor and BCGP predictor are 
comparable and about one half of that of the kriging predictors.  
In contrast, the uncertainty of the BCGP predictor as quantified by
the 95\% point-wise uncertainty bands is visually 
much smaller for the BCGP predictor than for any 
of the other three predictors. 
\ \ \ \   $\blacksquare$

\medskip

\noindent
{\bf Example~4.2} 
\citet{qian:06} \citep[see also][]{ba:12}
describe a computer simulation 
used in the design of a heat exchange system
for electronic cooling applications.
Denoted $y(\bm{x})$, the simulator output is 
the  total rate of steady state heat 
transfer from the source, 
in this case an electronic device,  
to a sink which dissipates the heat.  The $d = 4$ inputs to 
$y(\bm{x})$ are:
\begin{center}
{\singlespacing
\begin{tabular}{|c|l|c|c|}
\hline
\multicolumn{1}{|c|}{Notation} & 
\multicolumn{1}{|c|}{Description} &
\multicolumn{1}{|c|}{Lower Bnd} &
\multicolumn{1}{|c|}{Upper Bnd} \\
\hline
\hline
$x_1$ & Flow rate of entry air & 0.00055 & 0.001\\
$x_2$ & Temperature of entry air  & 270&  303.15\\
$x_3$ & Temperature of the heat source  & 330&  400\\
$x_4$ & Solid material thermal conductivity  &  202.4 & 360\\
\hline
\end{tabular}
}
\end{center}

\begin{figure}[t] 
\centering 
 \includegraphics[width=.6\linewidth]{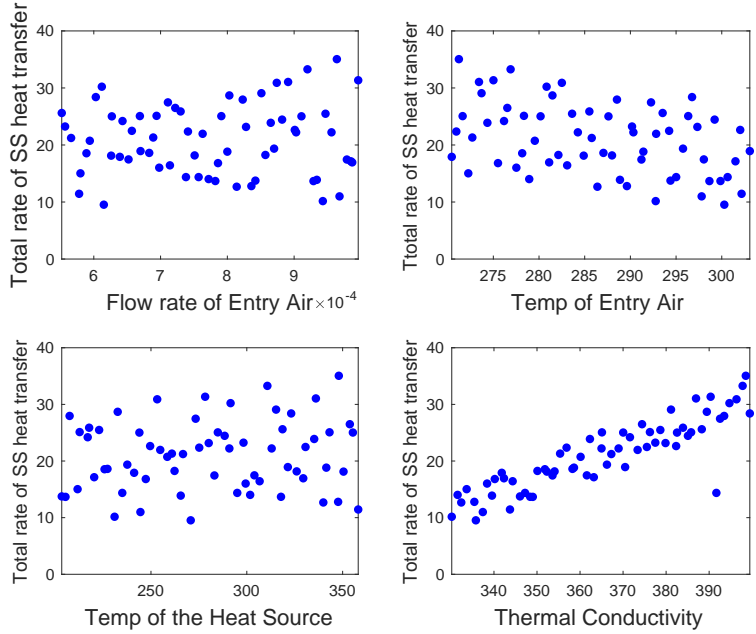}
\caption{\small \sl Marginal plots of state rate
of heat transfer versus $x_1$, $x_2$, $x_3$, $x_4$
for Example~4.2.}
\label{fig:HeatXchgeMarginalPlots}  
\end{figure}

\cite{qian:06} provide computed steady state
heat exchange values for 64 input vectors that form an
 orthogonal array-based Latin Hypercube design 
\citep[Chap.~5 of][]{SanWilNot2018}.
A design for the training data
containing 40 inputs from among the 64 available 
was selected to (approximately) maximize the minimum interpoint 
distance; the remaining 24 inputs were used 
as test data.  
As suggested by the marginal 
plots of the 
training  data shown in Figure~\ref{fig:HeatXchgeMarginalPlots},
$x_4$ appears to be the most active input influencing $y(\bm{x})$ 
while  $x_2$ also appears to be active but less so than
$x_4$.  Figure~\ref{fig:HeatXchgeMarginalPlots} also 
suggests $y(\bm{x})$ appears to be well modeled as a draw 
from a linear regression plus stationary deviation process.
This example will show that the BCGP model
can predict  $y(\bm{x})$ test data well
in stationary deviation cases such as this appears to be. 
  
As for Example~4.1, BCGP was 
run using 
$60,000 = 60 \times 1000$ iterations for calibration,
followed by 4,000 burn-in iterations, and 5,000 production
iterations; also, the $\omega$ prior was taken 
be the  $Beta(4, 6)$ distribution truncated to $[0.5, 1.0]$. 
The prior for $\rho_{G,j}$, $j=1,\ldots,4$, 
were taken to be independent and identically Beta(1.0,0.4) distributed.
The conditional distribution of $\rho_{L,j}$ given
$\rho_{G, j}$, $j=1,\ldots,4$, were taken to have independent Beta(1.0,1.0)
 distributions truncated to $[0, \rho_{G,j}]$.  

\begin{figure}[!h] 
\centering 
 \includegraphics[width=.85\linewidth]{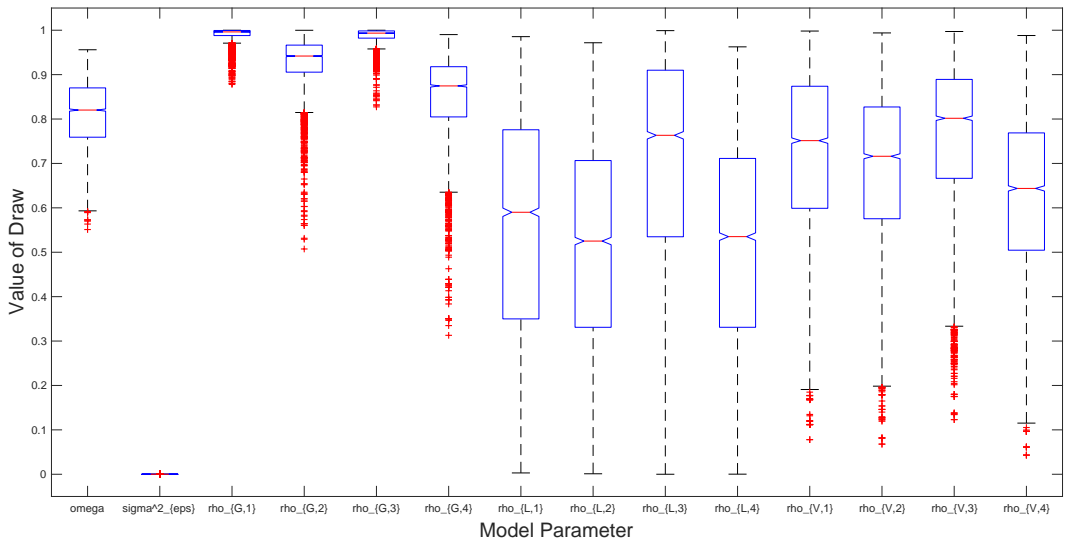}
\caption{\small \sl Boxplots of the posterior draws of all 
BCGP model parameters for Example~4.2}
\label{fig:PosteriorDraws}  
\end{figure}

Draws from the posterior distribution of the model parameters 
are shown in Figure~\ref{fig:PosteriorDraws}. 
The correlation parameters 
for the $Y_G(\bm{x})$ process, $\rho_{G,1}, \ldots,\rho_{G,4}$,
show that the inputs $x_2$ and $x_4$
appear most active because they have the smallest median draws, and 
the smaller $\rho_{G,4}$ values show that $x_4$ appears more active than $x_2$.
This is consistent with exploratory plots of the data in 
Figure~\ref{fig:HeatXchgeMarginalPlots}.  
Also judging by
the posterior correlation distributions, 
all four inputs are active for both the local deviation
process $Y_L(\bm{x})$ and the  $\sigma^2(\bm{x})$ process.  
Predicted surfaces for (the posterior means)
of any of these processes can be displayed for sections
of the input space that either fix 2 or more inputs
or traverse a curve in 4-space.   

\begin{figure}[h] 
\centering 
 \includegraphics[width=.45\linewidth]{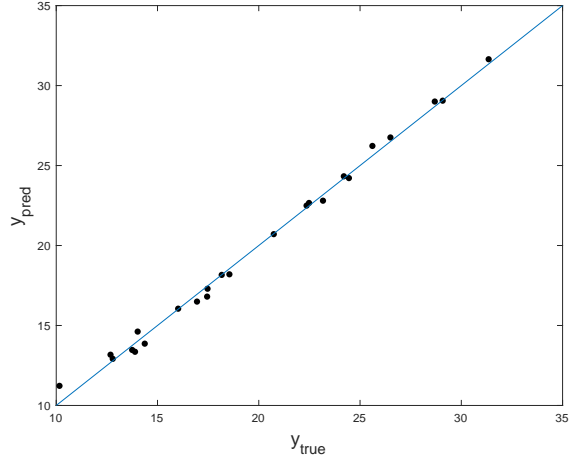}
\caption{\small \sl Predicted versus simulated values 
for the 24
steady state heat exchange inputs of Example~4.2}
\label{fig:Preds_of_24pts}  
\end{figure}

The posterior predictive mean of the $Y(\bm{x})$ process
was estimated at the 24 test data locations.
A plot of the simulated versus predicted values
is shown in Figure~\ref{fig:Preds_of_24pts}
overlaid with 
the $45^{\circ}$ line $y = x$.  
The predictions are, overall, extremely close to the
true values and the 
RMSPE of the 24 predictions is
$0.410$. The predictive accuracy is similar to that of CGP
(RMSPE equal to $0.438$) and the kriging predictor with a linear mean (RMSPE equal to $0.480$
when fit using REML with no nugget).  These RMSPEs are about half that of the  
kriging predictor having a constant mean (RMSPE equal to  $0.933$ when fit 
using REML with no nugget).
\ \ \ \   $\blacksquare$

\medskip

\noindent
{\bf Example~4.3} 
This final example 
contains a larger number of inputs, $d = 10$, than either Examples~4.1 or 4.2.
Additionally the output model is analytic so 
one can easily test the prediction
accuracy at arbitrary inputs.  
\citet{forr:08} state the equation  
\begin{equation}
\label{eq:wingweightsimulator}
y(\bm{x}) 
=
0.036 S_w^{0.758}
W_{wf}^{0.0035}
\left( \frac{A}{\cos^2(\Lambda)} \right)^{0.6}
q^{0.006}
\lambda^{0.04}
\left( \frac{100t_c}{\cos(\Lambda)} \right)^{-0.3}
\left( N_Z W_{dg}\right)^{0.49}  
+
S_w W_p
\end{equation}
for the 
{weight}
of a {light aircraft wing} as a function of the 10 
geometric/structural inputs with ranges provided
in Table~\ref{tab:ranges}.
Previous calculations of
the {\em total sensitivity indices}  
for this $y(\bm{x})$ have shown that
the most active inputs 
are, in order, $x_8 > x_3 > x_7 = x_1 > x_9$
and all other other inputs have only a minor impact.
\begin{table}[!h]
\begin{center}
{\singlespacing
\begin{tabular}{|c|l|c|}
\hline
Notation & 
\multicolumn{1}{|c|}{Input (units)} 
& Range \\
\hline
\hline
$x_1$/$S_w$ & wing area (ft$^2$) & $[150, 200]$ \\  
$x_2$/$W_{wf}$& weight of fuel in the wing (lb)&  $[220, 300]	$\\ 
$x_3$/$A$ & aspect ratio& $[6, 10]$ \\ 
$x_4$/$\Lambda$ & quarter-chord sweep (deg)& $[-10, 10]$ \\
$x_5$/$q$ &  dynamic pressure at cruise (lb/ft$^2$)& $[16, 45]$\\
$x_6$/$\lambda$  & taper ratio & $[0.5, 1]$\\
$x_7$/$t_c$ & aerofoil thickness to chord ratio & $[0.08, 0.18]$\\
$x_8$/$N_Z$ &  ultimate load factor& $[2.5, 6]$\\
$x_9$/$W_{dg}$ & flight design gross weight (lb)& $[1,700, 2,500]$\\
$x_{10}$/$W_p$ &   paint weight (lb/ft$^2$)&  $[0.025, 0.08]$\\
\hline
\end{tabular}
\caption{\small \sl Input variables and ranges for wing weight
in Example~4.3.
\label{tab:ranges}}
}
\end{center}
\end{table}

The BCGP predictor was  applied to $y(\bm{x})$
based on a 50 run input data set.
A $50 \times 10$ run maximin Latin hypercube design having 10 inputs
was selected as the input training data 
for predicting wing weight
(\url{https://spacefillingdesigns.nl}). 
Then a $150 \times 10$ matrix of test data inputs was formed using
 the Sobol\'{} sequence \citep[Chap.~5 of][]{SanWilNot2018}. 
Prediction and uncertainty quantification of $y(x)$
are based on the BCGP model with 
prior  as in Examples~4.1 and 4.2.   
In particular, the $\omega$ prior was the Beta$(4,6)$
distribution truncated to $[0.5, 1.0]$ while
 the prior for each of the $d=10$ global  correlations, 
 $\{ \rho_{G,j} \}_{j=1}^{10}$,
 were given independent Beta$(1,0.4)$ distributions.  
The MCMC sampling used
  $60,000 = 60 \times 1000$ 
calibration iterations followed by a 
larger numbers of burn-in (5,000) and production iterations (10,000)
for this larger $d$ example 
than for Examples~4.1 and 4.2.  

\begin{figure}[h] 
\centering 
\includegraphics[width = .4\linewidth]{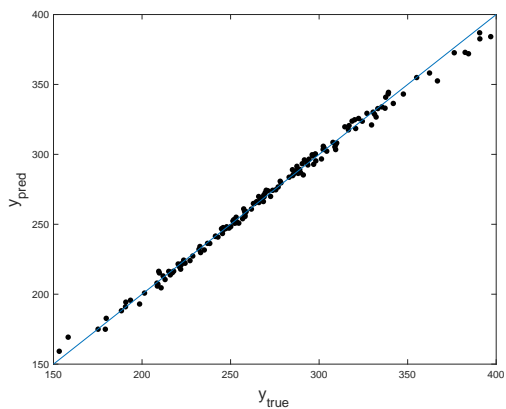}
\caption{\small \sl Predicted wing weight
versus calculated wing weight
for  150 test inputs based 
on 50 training inputs from a maximin LHD.
}
\label{fig:150predictions}  
\end{figure}
Figure~\ref{fig:150predictions} plots the 
150 predicted wing weights versus the simulated wing weights from 
(\ref{eq:wingweightsimulator}).  
The relative errors ranged from $1.1073\times 10^{-04}$
to  $0.0694$ and have mean $0.0097$.  
To compare the accuracy of the 
BCGP predictor with that of the CGP and two kriging
predictors, the RMSPE for the 150 test inputs was calculated.  
The RMSPE for the BCGP predictor was 3.62, for the CGP predictor
was 2.76, while that of the constant mean kriging predictor
was 1.03, and that of the linear mean kriging predictor was 0.91.
The kriging predictors are very accurate for this example.

\begin{figure}[h] 
\begin{center} 
\includegraphics[width = .4\linewidth]{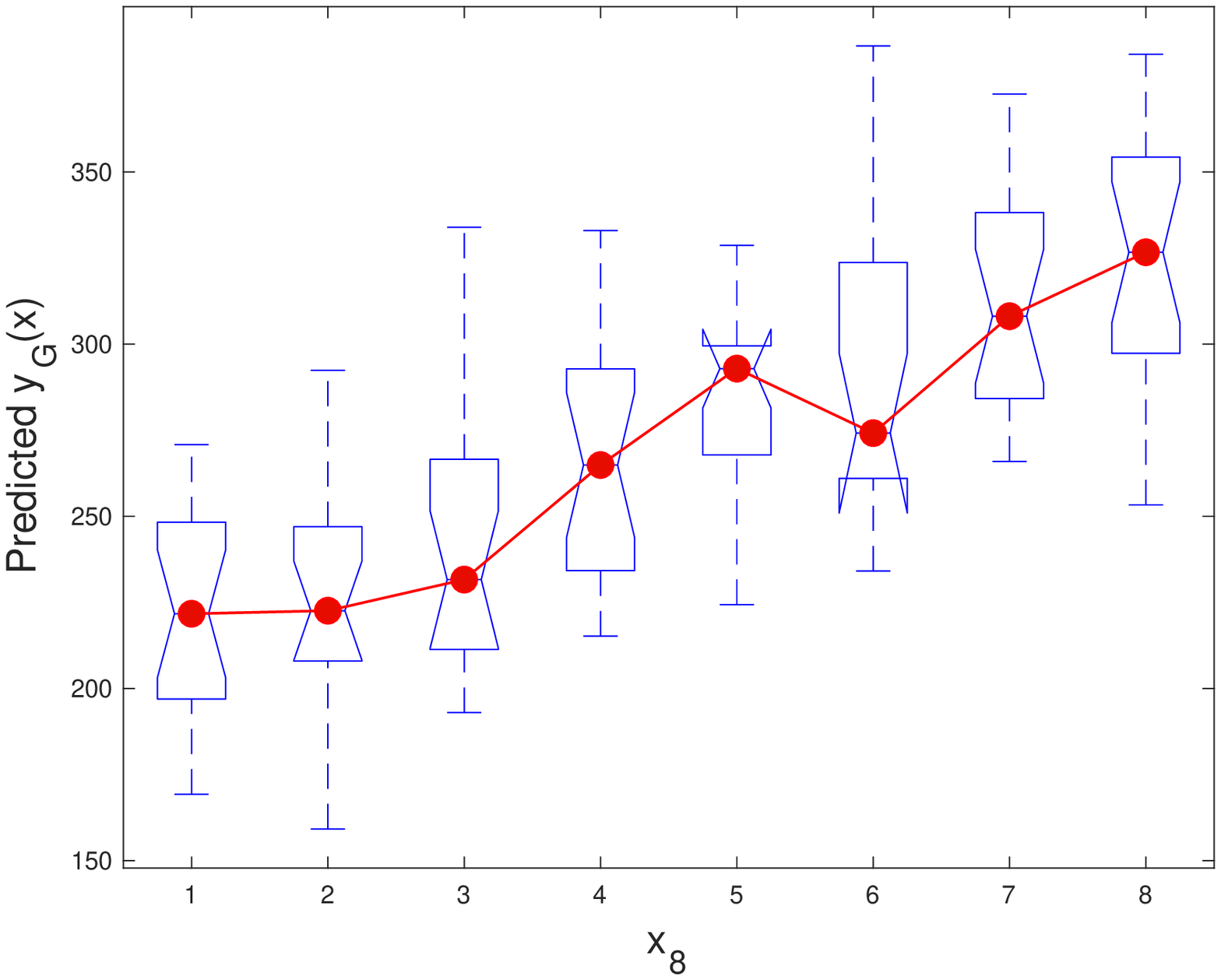}
\hspace{.1in}
\includegraphics[width = .4\linewidth]{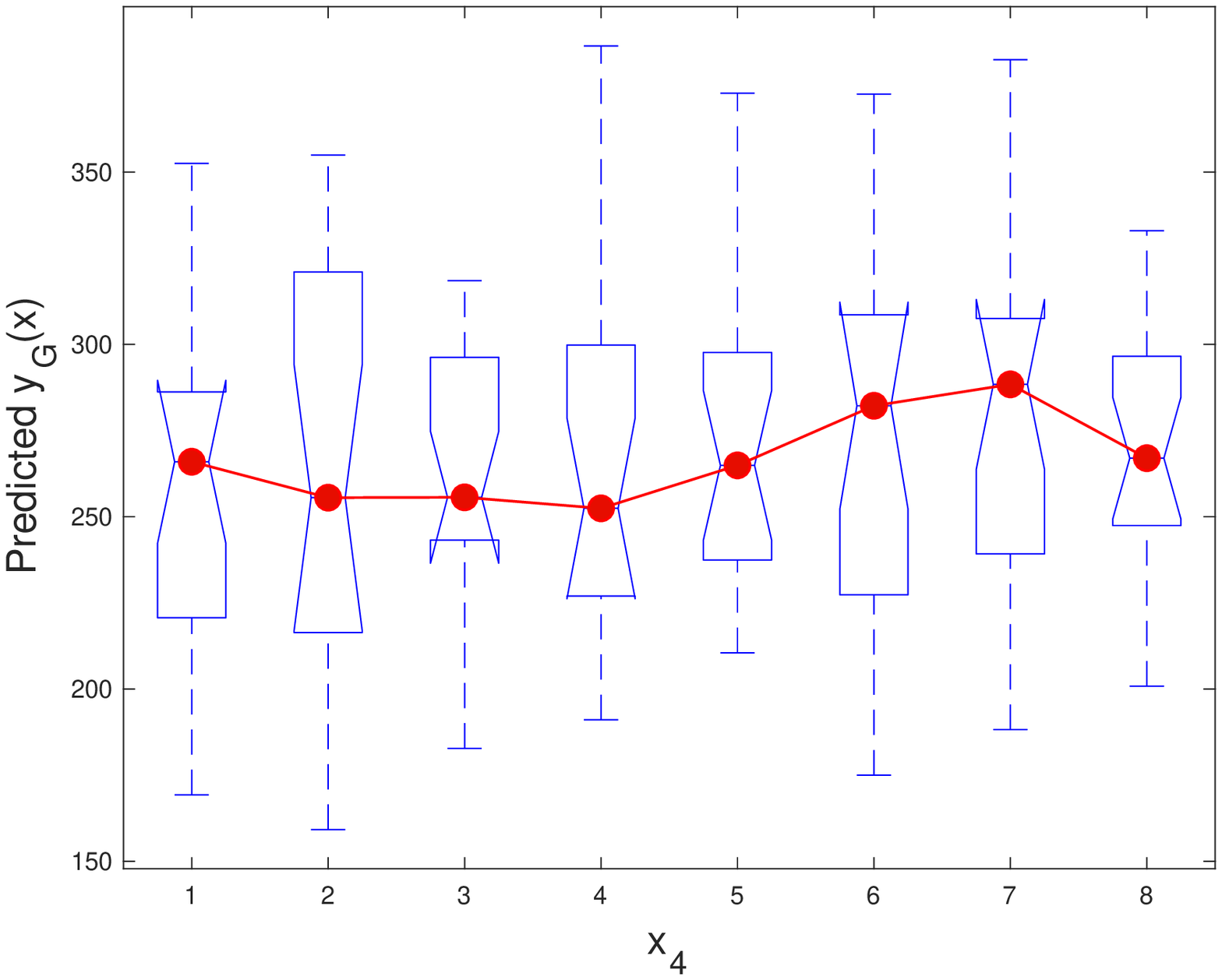}
\end{center} 
\vspace{-.3in}
\caption{\small \sl Boxplots of the predicted 
global trend function
for the wing weight function, $\widehat{y}_G(\bm{x})$, 
based on grouped  
$x_8$
and $x_4$ values for  150 test inputs. 
}
\label{fig:yG_predictions}  
\end{figure}

One opportunity that CGP and BCGP provide is the opportunity
to examine the  {\em global trend} 
curve, $\widehat{y}_G(\bm{x})$.  Here we consider
the activity of inputs on  $\widehat{y}_G(\bm{x})$. 
Recall that $x_8$, $x_3$, $x_9$,
were considered active for wing weight $y(\bm{x})$ while
$x_4$ was considered in-/low-activity. It is natural to speculate that
the same inputs are active or inactive for ${y}_G(\bm{x})$.
To examine this question, fix  $i \in \{1,\ldots,10\}$
and divide the range of  $x_i$ into 8 equal length subintervals.
Then group the 150 predicted $y_G(\bm{x})$ into 8 groups
according to the subinterval that $x_i$ for that prediction 
falls.
The grouped $y_G(\bm{x})$ predictions
are plotted as 8 side-by-side boxplots in Figure~\ref{fig:yG_predictions}.
Connecting the
medians of each boxplot shows that 
${y}_G(\bm{x})$ increases in $x_8$ while the analogous plot for 
$x_4$ showed low ${y}_G(\bm{x})$ activity.  Similar plots for 
$x_3$, $x_7$, $x_1$, and $x_9$ show these inputs to be 
active while those for 
$x_2$, $x_5$, and $x_6$  show low activity.
\ \ \ \   $\blacksquare$

\section{Summary and Discussion} 
\label{SummaryDiscussion}

This paper  proposes a Bayesian method to
predict the output from a computer simulator that produces 
possibly non-stationary output $y(\bm{x})$.
The methodology is developed to allow
output containing measurement error. 
 Based on a set of training data,  
prediction is based on a Bayesian Composite Gaussian process (BCGP) model
$Y(\bm{x})$. 
The BCGP is a hierarchical $Y(\bm{x})$ model that 
has the following features. 
The top stage of the BCGP model can be viewed 
as the sum of a global (mean) process, say  $Y_G(\bm{x})$,
and a local (deviation) process, say $Y_L(\bm{x})$.  
The global mean, say $y_G(\bm{x})$ which
is a draw from $Y_G(\bm{x})$, is meant to be a flexible description
of large-scale $y(\bm{x})$ trends. The local deviation,
say $y_L(\bm{x})$ which is a draw from $Y_L(\bm{x})$,  
captures small-scale $y(\bm{x})$ changes about $y_G(\bm{x})$.  
Subsequent stages put a prior 
on the global and local
process parameters that ensure $y_G(\bm{x})$ 
draws are
smoother than $y_L(\bm{x})$ draws.  
Another ingredient of the BCGP is that it contains
a model parameter which allows the data to 
weight the effect of the global 
and local processes.  
Lastly, the BCGP can describe $Y(\bm{x})$ 
having heteroscedastic process variability by using 
a latent variable process to describe
the  variance of $Y(\bm{x})$.
The method of prediction described in this paper allows one
to estimate the global and local components
of  $y(\bm{x})$.  The resulting predictions can be used, say, to determine
the sensitivity of $y_G(\bm{x})$ to each input.
Figure~\ref{fig:yG_predictions} of Example~4.3 
illustrates this approach.

One area for future research is refinement of the
prior, components of which have been selected
for their analytic tractability.  Most of our hyper-parameter 
choices have been made to reflect  vague prior information;
however, the choice of the
hyper-parameters $\alpha_\omega$ and $\beta_\omega$ 
for the $\omega$ prior are critical
in determining the properties of the
predicted $y_G(\bm{x})$ and $y_L(\bm{x})$.  
We have examined the global smoothness of 
the predicted $y_G(\bm{x})$,
the centeredness of the predicted 
$y_L(\bm{x})$ about zero, and their relative
smoothness 
for varying $\alpha_\omega$ and $\beta_\omega$.   
These properties were gauged heuristically
in a test series of analytic examples having known $y_G(\bm{x})$ 
which was perturbed by a $y_L(\bm{x})$ having low-activity
inputs.
 The final choice of parameters
for the $\omega$ prior was made based on the ability to 
predict the $y_G(\bm{x})$. 
Analysts in different subject matter
areas should do such an assessment using  test beds 
drawn from their applications.  This intuitive
method of selecting a prior is not the only option
for applying the prediction methodology introduced 
in this paper.  Two alternatives are the use 
of Reference Priors as described in \cite{GuWanBer2018}
and the prior used for the widely-used 
Bayesian calibration software
{\tt GMSPA} that is introduced in 
\citet{HigKenCav2004,HigGatWil2008} 
\citep[see also][]{Gat2008}.

The methods and priors described in 
this paper
are implemented in Matlab code 
that was used to run the examples in Section~\ref{sec:examples}.
This code is available from the first author.


\vspace{-.4in}
 \begin{center}
\item \subsection*{ACKNOWLEDGMENTS}
 \end{center}

 \vspace{-.1in}

 \doublespacing

This material was based upon work partially supported by the National Science Foundation under Grants DMS-0806134 and DMS-1310294
to The Ohio State University and under Grant DMS-1638521 to the Statistical and Applied Mathematical Sciences Institute. Any opinions, findings, 
and conclusions or recommendations expressed in this material are those of the author(s) and do not necessarily reflect the views of the National Science Foundation.\\


\end{document}